\documentclass[conference]{IEEEtran}
\IEEEoverridecommandlockouts
\usepackage[OT1]{fontenc}
\usepackage{amsmath,amssymb,amsfonts}
\usepackage{amsthm}
\usepackage{algorithmic}
\usepackage{graphicx}
\usepackage{textcomp}
\usepackage{xcolor}
\usepackage{xparse}
\usepackage{hyperref}
\usepackage{xspace}
\usepackage[most]{tcolorbox}
\usepackage{tabularray}
\usepackage{tabularx}
\usepackage{booktabs}
\usepackage{subcaption}
\usepackage[commandnameprefix=always]{changes}
\usepackage[switch]{lineno}
\usepackage[numbers,sort&compress]{natbib}
\usepackage{titlecaps}
\usepackage{listings}
\usepackage{lstlinebgrd}
\usepackage{tikz}
\usetikzlibrary{arrows.meta,positioning}
\usepackage{pgfplots}
\pgfplotsset{compat=1.18}

\usepackage{easyReview}
\usepackage{pifont}
\usepackage{microtype}

\newcounter{result}

\newtcolorbox{result}[1][]{
enhanced,
breakable,
use color stack,
boxrule=0.6pt,
arc=2pt,
colback=blue!2!white,
colframe=blue!55!black,
borderline west={2pt}{0pt}{blue!55!black}, left=3pt,
right=3pt,
top=2pt,
bottom=2pt,
before skip=3pt,
after skip=3pt,
fontupper=\normalfont,
before upper={
    \refstepcounter{result}\textbf{Result~\theresult:}\quad
},
#1
}

\newcounter{finding}

\newtcolorbox{finding}[1][]{
enhanced,
breakable,
use color stack,
boxrule=0.6pt,
arc=2pt,
colback=teal!3!white,
colframe=teal!60!black,
borderline west={2pt}{0pt}{teal!60!black}, left=3pt,
right=3pt,
top=2pt,
bottom=2pt,
before skip=3pt,
after skip=3pt,
fontupper=\normalfont,
before upper={
    \refstepcounter{finding}\textbf{Finding~\thefinding:}\quad
},
#1
}

\newtheoremstyle{rccdefinitionstyle}
{0pt}
{0pt}
{\normalfont}
{}
{\color{brown!80!black}\bfseries}
{.}
{0.5em}
{}
\theoremstyle{rccdefinitionstyle}
\newtheorem{definition}{Definition}
\tcolorboxenvironment{definition}{
enhanced,
breakable,
use color stack,
colback=black!5,
frame hidden,
boxrule=0pt,
arc=0mm,
left=2.4mm,right=2mm,top=1.2mm,bottom=1.2mm,
borderline west={1.8pt}{0pt}{brown!80!black},
before skip=8pt, after skip=8pt
}

\usepackage{enumitem}
\setenumerate[1]{itemsep=0pt,partopsep=0pt,parsep=\parskip,topsep=2pt,leftmargin=12pt}
\setitemize[1]{itemsep=0pt,partopsep=0pt,parsep=\parskip,topsep=2pt,leftmargin=12pt}

\newcommand\rcc{{\sf rcc}}

\newcommand\Dd{{\mathcal{D} }}
\newcommand\Pp{{\mathcal{P} }}
\newcommand\Kk{{\mathcal{K} }}

\newcommand\Rr{{\mathcal{R} }}

\NewDocumentCommand{\bench}{g}{ChainVul\IfValueT{#1}{\textsubscript{#1}}\xspace
}

\newcommand\TP{{\sf TP}}
\newcommand\TN{{\sf TN}}
\newcommand\FP{{\sf FP}}
\newcommand\FN{{\sf FN}}

\newcommand\DTP{{\Dd_\TP}}
\newcommand\DTN{{\Dd_\TN}}
\newcommand\DFP{{\Dd_\FP}}
\newcommand\DFN{{\Dd_\FN}}

\newcommand\lam{\lambda^\text{m}}
\newcommand\lan{\lambda^\text{n}}
\newcommand\funcn{{\texttt{nntp()}}}
\newcommand\funcm{{\texttt{mutt()}}}
\newcommand\fcmsg{{\texttt{fc.messages}}}

\newcommand\citesys{\cite{openai, deepseek, qwen35_27b, opencode, swe-agent}}

\newcommand\citerw{{\cite{Big-Vul, primevul, JitVul, correct, VulnSage, svd-bench, comfortzone, VulDetectBench, SecVulEval}}}

\newcommand\citeev{\cite{MultiTask, AdditionalInfo, primevul, JitVul, correct, VulnSage, svd-bench, comfortzone, ts-vuls, VulDetectBench, SecVulEval}}

\newcommand\citebe{\cite{Big-Vul, primevul, svd-bench, JitVul, VulnSage}}
\newcommand\citebl{\cite{VulDetectBench, SecVulEval}}
\newcommand\citebc{\cite{VulDetectBench}}
\newcommand\citebr{\cite{correct, SecVulEval}}

\newcounter{rqgroup}
\newcounter{rq}[rqgroup]

\renewcommand{\therq}{RQ\arabic{rqgroup}.\arabic{rq}}

\newcommand{\RQGroup}[1]{\setcounter{rqgroup}{#1}\setcounter{rq}{0}}

\newcommand{\RQ}[2]{\refstepcounter{rq}\label{#1}\expandafter\gdef\csname rq@text@#1\endcsname{#2}\textbf{\therq:}~(\textbf{#2})}

\newcommand{\rqref}[1]{\ref{#1}}

\newcommand{\rqfullref}[1]{\ref{#1}:~\csname rq@text@#1\endcsname
}

\newcommand{\rqtext}[1]{\csname rq@text@#1\endcsname
}

\begin{document}
\title{
ChainVul: A Root-Cause-Chain-Based Vulnerability Benchmark for Precise LLM and Agent Assessment
}
\newcommand{\email}[1]{#1}
\author{\setlength{\parindent}{0pt}\begin{minipage}{\dimexpr\textwidth-20pt\relax}\centering
  \setlength{\tabcolsep}{2pt}\begin{tabular}{@{}*{4}{c}@{}}
    \parbox[t]{5.4cm}{\centering
      \sublargesize Tiantian Sun\\ \small\textit{School of Advanced Interdisciplinary\\ Sciences, University of Chinese\\ Academy of Sciences}\\
      \small Beijing, China\\
      \small\textit{Institute of Software,\\ Chinese Academy of Sciences}\\
      \small Beijing, China\\ {\small\email{suntiantian24@otcaix.iscas.ac.cn}}} &
    \parbox[t]{3.6cm}{\centering
      \sublargesize Yuyang Xie\\ \small\textit{Beijing Jiaotong University}\\
      \small Beijing, China\\ {\small\email{xieyuyang@bjtu.edu.cn}}} &
    \parbox[t]{4.0cm}{\centering
      \sublargesize Jinlong He\\ \small\textit{Institute of Software, Chinese Academy of Sciences}\\
      \small Beijing, China\\ {\small\email{hejinlong@otcaix.iscas.ac.cn}}} &
    \parbox[t]{4.0cm}{\centering
      \sublargesize Shixin Zhang\\ \small\textit{Institute of Software, Chinese Academy of Sciences}\\
      \small Beijing, China\\ {\small\email{zhangshixin@otcaix.iscas.ac.cn}}} \\
  \end{tabular}
  \par\vspace{4ex}
  \setlength{\tabcolsep}{12pt}\begin{tabular}{@{}*{3}{c}@{}}
    \parbox[t]{4.9cm}{\centering
      \sublargesize Jiwei Yan\textsuperscript{*$\dagger$}\\ \small\textit{Beijing Key Laboratory of Intelligent Software Engineering}\\
      \small Beijing, China\\
      \small\textit{Institute of Software, Chinese Academy of Sciences}\\
      \small Beijing, China\\ {\small\email{yanjiwei@otcaix.iscas.ac.cn}}} &
    \parbox[t]{4.9cm}{\centering
      \sublargesize Jun Yan\textsuperscript{*$\dagger$}\\ \small\textit{Beijing Key Laboratory of Intelligent Software Engineering}\\
      \small Beijing, China\\
      \small\textit{Key Laboratory of System Software (Chinese Academy of Sciences)}\\
      \small\textit{Institute of Software, Chinese Academy of Sciences}\\
      \small Beijing, China\\ {\small\email{yanjun@ios.ac.cn}}} &
    \parbox[t]{4.9cm}{\centering
      \sublargesize Jian Zhang\textsuperscript{$\dagger$}\\ \small\textit{Key Laboratory of System Software (Chinese Academy of Sciences)}\\
      \small\textit{Institute of Software, Chinese Academy of Sciences}\\
      \small Beijing, China\\ {\small\email{zj@ios.ac.cn}}} \\
  \end{tabular}
  \end{minipage}
  \thanks{\textsuperscript{*}Corresponding authors.}
  \thanks{\textsuperscript{$\dagger$}Also with University of Chinese Academy of Sciences.}
}
\maketitle

\begin{abstract}
Large language models (LLMs) and agents are increasingly adopted for vulnerability detection, yet precisely evaluating their effectiveness on real-world vulnerabilities remains challenging. Existing benchmarks suffer from two major limitations: (1) unclear boundaries of complete vulnerability-relevant code contexts, and (2) coarse-grained metrics for judging detection results. These limitations hinder the reliable evaluation of vulnerability detection results produced by LLMs and agents.
To address these issues, we propose an LLM- and agent-oriented approach to constructing vulnerability benchmarks and build \textit{ChainVul}, which comprises 1,000 vulnerability samples. The core design of \textit{ChainVul} is the extraction of Root Cause Chain (RCC), a novel representation for capturing the root cause of vulnerabilities. Based on RCC, \textit{ChainVul} enables more precise evaluation through both RCC-oriented code context identification and fine-grained RCC-based assessment metrics.

Our evaluation demonstrates the effectiveness of this design.  In \textit{ChainVul}, nearly 60\% of the samples involve complex vulnerability contexts that cannot be adequately characterized by conventional patch-based annotations alone, highlighting the necessity of explicit RCC modeling. 
{Through manual inspection, we find that the extracted RCCs achieve 87.8\% key-statement coverage and 98.0\% triggering-condition accuracy. We also observe that RCC-based matching improves evaluation accuracy by around 30\%. }
Finally, we evaluate four LLMs and two agents on \textit{ChainVul}, providing deeper insights into their strengths and limitations in real-world vulnerability detection scenarios.

\end{abstract}


\section{Introduction}

Recent advances in large language models (LLMs) have triggered an explosive increase in software code generation, making software reliability and security more critical than ever. To improve software reliability and security in this context, a growing number of LLMs and agents~{\citesys} (hereafter referred to as LLM-based systems) are adopted for vulnerability detection.
However, despite their growing popularity, the extent to which these systems are effective for real-world vulnerability detection remains unclear. This, in turn, necessitates a high-quality \textit{vulnerability benchmark} that integrates both representative \textit{datasets} and practical \textit{evaluation framework} to enable accurate and fair assessment of LLM-based systems.

Due to the critical need for evaluating the vulnerability detection capabilities of LLM-based systems, numerous new benchmarks have been proposed, while existing traditional ones have also been reused for assessment.
According to the provenance of vulnerable code, these benchmarks can be categorized into \textit{manually constructed} vulnerability benchmarks~{\cite{juliet, castle, QuixBugs, Ghera, FlaskVuln, Vulcorpus}} and \textit{real-world} ones~{\cite{Big-Vul, primevul, CrossVul, diverseVul, Devign, ExcePy, Vul4J, PatchDB, CWEVuln, D2A}}.
The manually constructed benchmarks typically consist of artificially introduced vulnerabilities in controlled environments, which are relatively easy to build and provide clear vulnerability boundaries. 
However, due to their simplicity, modern LLM-based systems often achieve saturated performance, making it difficult to differentiate the true capabilities among different systems.
In contrast, vulnerabilities in real-world benchmarks are collected from actual software issues or CVE (Common Vulnerabilities and Exposures) reports, providing more realistic and diverse vulnerability samples for evaluating LLM-based systems. Therefore, in this paper, we will focus on constructing a high-quality \textit{real-world vulnerability benchmark}.

To further motivate this design, we conducted a systematic analysis of real-world vulnerability benchmarks~{\citerw} used in existing studies and observed that \textbf{many benchmarks used in evaluating LLM-based systems still inherit design principles originally developed for evaluating static application security testing (SAST) tools}. 
In fact, current evaluations mainly rely on the following two assumptions, which are inherited from SAST evaluation practices.
\textbf{Assumption \ding{172}:} the tools under evaluation can take the entire code repositories as analysis input. Thus, they do not require benchmarks to provide fine-grained contextual information about vulnerable code, i.e., only providing patch-related lines or the functions containing them for each vulnerability is enough for further evaluation.
\textbf{Assumption \ding{173}:} the tools under evaluation employ deterministic detection logic and seldom report arbitrary issues. Thus, adopting relaxed matching criteria to evaluate is acceptable.
Accordingly, a detection is considered successful as long as it is reported near the ground-truth vulnerable location.

However, we find that both assumptions break down for LLM-based systems, based on two key observations.
\textbf{Observation \ding{172}:} LLM-based approaches cannot effectively process entire repositories and instead rely on limited code contexts.
Consequently, benchmarks need to provide more complete vulnerability-related contextual information to enable accurate reasoning and detection.
\textbf{Observation \ding{173}:} LLM-based systems are not governed by deterministic detection rules and may produce reports based on arbitrary or incomplete reasoning. Therefore, stricter matching criteria are required to ensure reliable evaluation results.
That means, evaluating LLM-based systems puts higher requirements on the design of benchmarks.

Based on these observations, we derive the following key insight: \textbf{accurate evaluation of LLM-based systems requires fine-grained vulnerability annotations from both the \textit{code input} and \textit{bug report} matching perspectives}. In particular, such evaluation demands detailed information about each vulnerability, which necessitates uncovering its root cause.
Motivated by this insight, we propose an LLM- and agent-oriented vulnerability benchmark for precise evaluation, which consists of both a dataset and an evaluation framework. 
At the core of our benchmark, we introduce a new structured representation, termed \textbf{Root Cause Chain (RCC)}, which captures the root-cause reasoning-related elements (\textit{vulnerability-relevant code statements} and \textit{vulnerability-triggering key conditions}) underlying vulnerability manifestation.
Based on RCC, we construct an \textit{RCC-based dataset} where each vulnerability sample is associated with an RCC annotation and the relevant vulnerability context. In addition, we design an \textit{RCC-aware evaluation framework} that enables fine-grained assessment of whether an LLM correctly identifies the target vulnerability. Together, these two components constitute the \textit{RCC-based benchmark}, \textbf{ChainVul}. In our implementation, \textit{ChainVul} contains 1,000 vulnerability samples. 

{In the evaluation, we first validate the quality of the RCC annotations via manual inspection, which shows that RCC achieves 87.8\% key-statement coverage and 98.0\% triggering-condition accuracy. We further show that, compared with static-analysis-based context construction methods, RCC-based context achieves the highest 97.4\% key-statement coverage with a compact context (2.2k tokens median).} RCC-based context also consistently improves the performance of four LLMs, yielding gains of up to 13\% in F1-score and 16\% in paired accuracy. These results highlight the necessity of explicit root-cause modeling for vulnerability evaluation. Moreover, RCC-based matching improves evaluation accuracy by around 30\%, demonstrating the importance of fine-grained, root-cause-aware assessment criteria. Finally, we evaluate four LLMs and two agents on \textit{ChainVul},
revealing that both context completeness and context utilization substantially affect vulnerability detection performance.

The contributions of this paper are as follows:
\begin{itemize}
\item We introduce \textit{Root Cause Chain (RCC)}, a novel representation that captures both the \textit{vulnerability-relevant code statements} and \textit{vulnerability-triggering key conditions} underlying vulnerability manifestation, providing a foundation for precise evaluation of LLM-based detection systems.

\item We propose an RCC-based benchmark construction approach and build \textit{ChainVul}, an LLM- and agent-oriented vulnerability benchmark consisting of 1,000 vulnerability samples, each precisely annotated with an RCC. Both our dataset and code are publicly available~\cite{ourtool}.

\item We conduct comprehensive experiments on six LLM- and agent-based systems using \textit{ChainVul}. The results reveal how context completeness and fine-grained evaluation criteria influence vulnerability detection performance, providing new insights into the strengths and limitations of existing systems in real-world scenarios.

\end{itemize}

\lstset{
	basicstyle=\ttfamily\scriptsize,
	numbers=left,
	numberstyle=\tiny\color{gray!80},
	numbersep=6pt,
	xleftmargin=1.2em,
	frame=single,
	breaklines=true,
	columns=fullflexible,
	tabsize=2,
	keepspaces=true,
	showstringspaces=false,
	keywordstyle=\color{blue!70!black}\bfseries,
	commentstyle=\color{green!40!black}\itshape,
	stringstyle=\color{orange!80!black},
	rulecolor=\color{black!35},
	backgroundcolor=\color{black!1}
}

\begin{figure*}[ht]
\centering
\begin{subfigure}[b]{0.45\linewidth}
\begin{lstlisting}[language=C,
linebackgroundcolor={\ifnum\value{lstnumber}=6 \color{green!18}\fi
\ifnum\value{lstnumber}=7 \color{green!18}\fi
}]
static int nntp_fetch_headers(...) {
	struct NntpData *nntp_data = ctx->data;
	struct FetchCtx fc;
	char buf[HUGE_STRING];
	fc.messages = mutt_mem_calloc(length, 1);
+	if (fc.messages == NULL)
+	   return -1;
	if (rc == 0) {
		if (!fc.messages[idx])
			snprintf(buf, sizeof(buf), "%u", current);
	}
	snprintf(..., fc.messages[idx], nntp_data->group);
}
\end{lstlisting}
\centering
\caption{File: nntp.c}
\label{fig:motivating_example_nntp}
\end{subfigure}
\hfill
\begin{subfigure}[b]{0.19\linewidth}
\begin{lstlisting}[language=C]
void *mutt_mem_calloc(
  size_t nmemb,
  size_t size
) {
	void *p = NULL;
	if (!nmemb || !size)
		return NULL;
	p = calloc(nmemb, size);
	if (!p)
		mutt_exit(1);
	return p;
}
\end{lstlisting}
\centering
\caption{File: memory.c}
\label{fig:motivating_example_memory}
\end{subfigure}
\hfill
\centering
\begin{subfigure}[b]{0.30\linewidth}
\begin{tikzpicture}[
	>=Latex,
	vnode/.style={circle, draw, minimum size=7mm, inner sep=0pt, font=\small},
	key/.style={vnode, fill=red!12},
	sup/.style={vnode, fill=blue!8}
]
\node[vnode] (N2) at (0,4.5) {$\lan_{2}$};
\node[sup]  (N5) at (0,3.2) {$\lan_{5}$};
\node[sup]  (N8) at (0,1.9) {$\lan_{8}$};
\node[key]  (N9) at (1.4,0.6) {$\lan_{9}$};
\node[key]  (N12) at (0,0.6) {$\lan_{12}$};
\node[sup]  (M6) at (2.7,3.2) {$\lam_{6}$};
\node[key]  (M7) at (2.7,1.9) {$\lam_{7}$};

\node[sup, minimum size=4mm] (LS) at (0.75,4.55) {};
\node[key, minimum size=4mm] (LK) at (0.75,4.05) {};
\draw[->, thick, black] (0.4,5.0) -- (0.95,5.0);
\node[anchor=west, font=\scriptsize] at (1.05,5.0) {execution path};
\node[anchor=west, font=\scriptsize] at (1.05,4.55) {supporting statement};
\node[anchor=west, font=\scriptsize] at (1.05,4.05) {key statement};

\draw[->, thick, red!80!black] (N2) to[bend right=30] (N5);
\draw[->, thick, red!80!black] (N5) to[bend left=20] (M6);
\draw[->, thick, red!80!black] (M6) to[bend left=20] (M7);
\draw[->, thick, red!80!black] (M7) to[bend left=20] (N5);
\draw[->, thick, red!80!black] (N5) -- (N8);
\draw[->, thick, red!80!black] (N8) -- (N9) node[midway, above right, xshift=1pt] {$p_1$};

\draw[->, thick, orange!90!black] (N2) to[bend left=30] (N5);
\draw[->, thick, orange!90!black] (N5) to[bend left=35] (M6);
\draw[->, thick, orange!90!black] (M6) to[bend left=35] (M7);
\draw[->, thick, orange!90!black] (M7) to[bend left=35] (N5);
\draw[->, thick, orange!90!black] (N5.west) .. controls (-1.0,3.2) and (-1.0,1.1) .. (N12.north west) node[pos=0.55, left] {$p_2$};

\draw[->, thick, green!60!black] (N2.west) .. controls (-1.8,4.5) and (-1.8,0.9) .. (N12.west) node[pos=0.5, left, xshift=-2pt] {$p'$};
\end{tikzpicture}
\centering
\caption{{Execution Paths}}
\label{fig:motivating_example_paths}
\end{subfigure}

\caption{Motivating example with two source snippets and execution-path graph.}
\label{fig:motivating_example}
\vspace{-12pt}
\end{figure*}
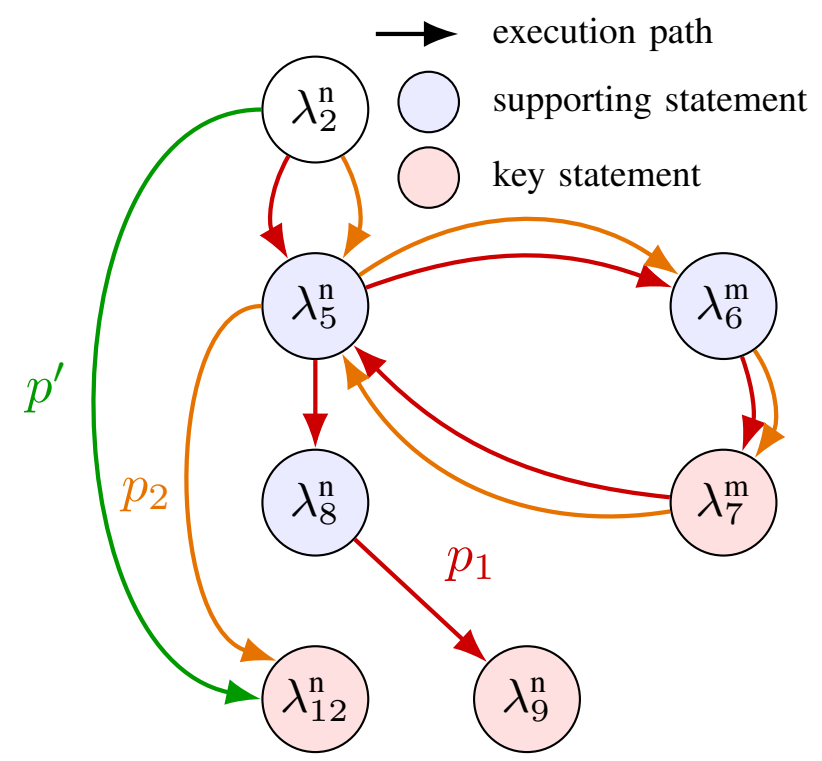

\section{Existing Benchmarks and Motivating Example}

In this section, we introduce the benchmarks used in existing evaluation studies and illustrate the challenges they face through a representative motivating example.

\subsection{Benchmarks Used in LLM-based System Evaluation}

In recent studies~{\citeev} that evaluate the vulnerability detection capabilities of LLM-based systems, many real-world benchmarks~{\citerw} have been adopted.
To understand their suitability in such evaluations, we focus on two dimensions: how they provide vulnerability-related code contexts within a repository, and which metrics they employ to determine whether a vulnerability has been detected.

The first dimension concerns the types of code contexts provided to LLM-based systems in existing benchmarks. Because these systems perform vulnerability analysis based exclusively on the supplied code context, the completeness of the context plays a critical role in determining whether vulnerabilities can be correctly identified. There are two commonly used types.

\begin{itemize}
    \item \textbf{Unexpanded Patch Context.}
This type of context directly adopts the functions modified in the fixing patch, without expanding other functions relevant to the vulnerability. These benchmarks either take each modified function as an independent sample~\cite{Big-Vul, primevul, svd-bench, comfortzone}, or take the set of modified functions in a fixing commit as a sample~\cite{VulnSage}.

\item \textbf{Rule-based Expanded Context.}
This type of context contains extra functions expanded from functions modified in the fixing patch with predefined rules, including location-expand rule~\cite{ts-vuls}, call-relation-expand rule~\cite{JitVul, correct}, and def-relation-expand rule~\cite{correct}, etc. 
Since these rules do not analyze the vulnerability, but rather consider the structure or call relationships, they are syntactic-level expansions.
\end{itemize}

The second dimension concerns the evaluation metrics adopted by existing benchmarks, which can be broadly categorized into four types. These metrics determine the criteria for judging whether the reports generated by LLM-based vulnerability detection systems correspond to real vulnerabilities. Since they impose different levels of strictness, the same detection result may be considered correct under one metric but incorrect under another, potentially leading to different conclusions regarding model performance.

\begin{itemize}
\item \textbf{Existence-aware Metric.}
This type of metric considers a vulnerability to be successfully detected as long as the generated report indicates the existence of the target vulnerability, regardless of its specific location or root cause. As the least restrictive evaluation criterion, it has been adopted by many existing benchmarks~{\citebe}.
\item \textbf{Location-aware Metric.}
This type of metric evaluates whether the generated report correctly localizes the target vulnerability. A report is regarded as correct only if the identified location is consistent with the ground truth, such as the corresponding file, function, or statement. This metric is more stringent than the existence-aware metric and has been adopted by several existing benchmarks~{\citebl}.
\item \textbf{Category-aware Metric.}
This type of metric evaluates whether the reported vulnerability belongs to the expected vulnerability category. Therefore, each sample in the benchmarks~{\citebc} is required to specify the vulnerability type, typically using CWE (Common Weakness Enumeration) identifiers. Compared with existence- and location-aware metrics, this criterion imposes additional semantic requirements on vulnerability identification.
\item \textbf{Reasoning-aware Metric.}
This type of metric evaluates the underlying reasoning process that leads to the generation of the vulnerability reports. A report is considered correct only if its reasoning process is fully consistent with the annotated ground-truth reasoning process~{\citebr}. While it provides strong semantic supervision, the requirement of exact reasoning matching may be overly strict, since different LLM-based systems can arrive at the same root cause through alternative yet logically valid reasoning paths.
\end{itemize}

\subsection{Motivating Example: A Vulnerability Sample}

In this subsection, we present a simplified yet real-world vulnerability sample collected from the benchmark \textit{JitVul}~\cite{JitVul}, which is shown in Fig.~\ref{fig:motivating_example_nntp} and \ref{fig:motivating_example_memory}.
Two code files (i.e., \texttt{nntp.c} and \texttt{memory.c}) contain two vulnerability-related functions: \texttt{nntp\_fetch\_headers} (referred to as \funcn) and \texttt{mutt\_mem\_calloc} (referred to as \funcm). For a convenient representation of code lines, we denote the $i$-th line in \texttt{nntp.c} and \texttt{memory.c} as $\lan_i$ and $\lam_i$, respectively. Specifically, the fixing patch introduces two additional lines in \texttt{nntp.c} ($\lan_6$ and $\lan_7$).

The {null pointer dereference (NPD)} bug is triggered in the following process: At $\lan_5$, the function {\funcm} is called to allocate a block of memory for {\fcmsg}. In the callee function {\funcm}, the arguments \texttt{nmemb} and \texttt{size} are checked at $\lam_6$, if either of them is zero, the function returns \texttt{NULL} ($\lam_7$).
The return value is then set to {\fcmsg} ($\lan_5$), which creates a hidden danger of NPD. In the subsequent code of {\funcn}, any code that dereferences {\fcmsg} triggers an NPD vulnerability.
At $\lan_9$, {\fcmsg} is dereferenced to check an element of the message array, an NPD could trigger here; at $\lan_{12}$, {\fcmsg} is dereferenced to set an element of the message array, an NPD could be triggered here too. This process describes a vulnerability that can be triggered by two concrete execution paths, as shown in Fig.~\ref{fig:motivating_example_paths}.
\begin{itemize}
    \item {$p_1$(red): $\lan_2 \to \lan_5 \to \lam_6 \to \lam_7 \to \lan_5 \to \lan_8 \to \lan_{9}$}
    \item {$p_2$(orange): $\lan_2 \to \lan_5 \to \lam_6 \to \lam_7 \to \lan_5 \to \lan_{12}$}
\end{itemize}

In addition, this example contains another NPD that differs from the target vulnerability. Its triggering process is as follows: At $\lan_2$, the \texttt{nntp\_data} is initialized with \texttt{ctx->data}, and the variable is then dereferenced at $\lan_{12}$. The execution path $p': \lan_2\to\lan_{12}$ triggers an NPD when \texttt{ctx->data} is a null pointer, as shown in Fig.~\ref{fig:motivating_example_paths} in green.

\subsection{Limitations in Using Existing Benchmarks}

The evaluation of LLM-based systems is not only a matter of collecting real-world vulnerabilities, but also of deciding what code context to provide and what criterion to use when matching reports to ground truth.
Based on the above example, we will further discuss the limitations of using existing benchmarks to evaluate the capability of LLM-based systems in analyzing such vulnerabilities.

\textbf{$\blacktriangleright$ Taking insufficient or noisy code context as evaluation input.} From the perspective of evaluation inputs, existing context-providing strategies may result in insufficient or noisy code context and consequently lead to false negatives (FN) or false positives (FP) in the evaluation results.
\label{sec:limitation-context}

\textit{Insufficient context.} Using unexpanded patch context may lead to incomplete semantic information. In the motivating example, the target vulnerability depends on both {\funcn} and {\funcm}. However, the unexpanded patch context includes only {\funcn}, preventing the system from inferring that the allocation at $\lan_5$ may return \texttt{NULL}, and thus potentially missing the inter-procedural NPD.

\textit{Noisy context.}
The opposite problem occurs when benchmarks expand context with generic rules. In this example, {\funcn} has 67 direct callers and callees, while the truly relevant context spans only two functions. Because these expansions follow syntactic relations rather than the semantics of vulnerability, they may introduce large amounts of unrelated code, dilute the vulnerable signal, and distort evaluation.

\textbf{$\blacktriangleright$ Using coarse-grained metrics to judge detection results.}
\label{sec:limitation-metric}
Besides providing the right context for precise evaluation, it also requires a high-quality metric to distinguish whether a reported issue matches the target vulnerability. In particular, a correct report should correspond to the same vulnerability that the patch fixes, not merely share a location or category.

We also use the motivating example to illustrate this issue. In this case, $p_1$ and $p_2$ correspond to the target vulnerability, whereas $p'$ represents a different NPD caused by dereferencing another variable. Under existence-aware, location-aware, or category-aware evaluation metrics, reporting $p'$ may be incorrectly considered as successfully detecting the target vulnerability, thereby leading to FP evaluation results.
Conversely, reasoning-aware metrics may introduce the opposite issue. The same vulnerability can be correctly identified through different valid analysis paths, such as forward tracing from $\lam_7$ or backward tracing from $\lan_9$. However, treating only one trace as correct would incorrectly reject other valid analyses, leading to FN judgments. As a result, these metrics may also mischaracterize detection performance.

\begin{figure*}[t]
    \centering
    \includegraphics[width=0.95\textwidth]{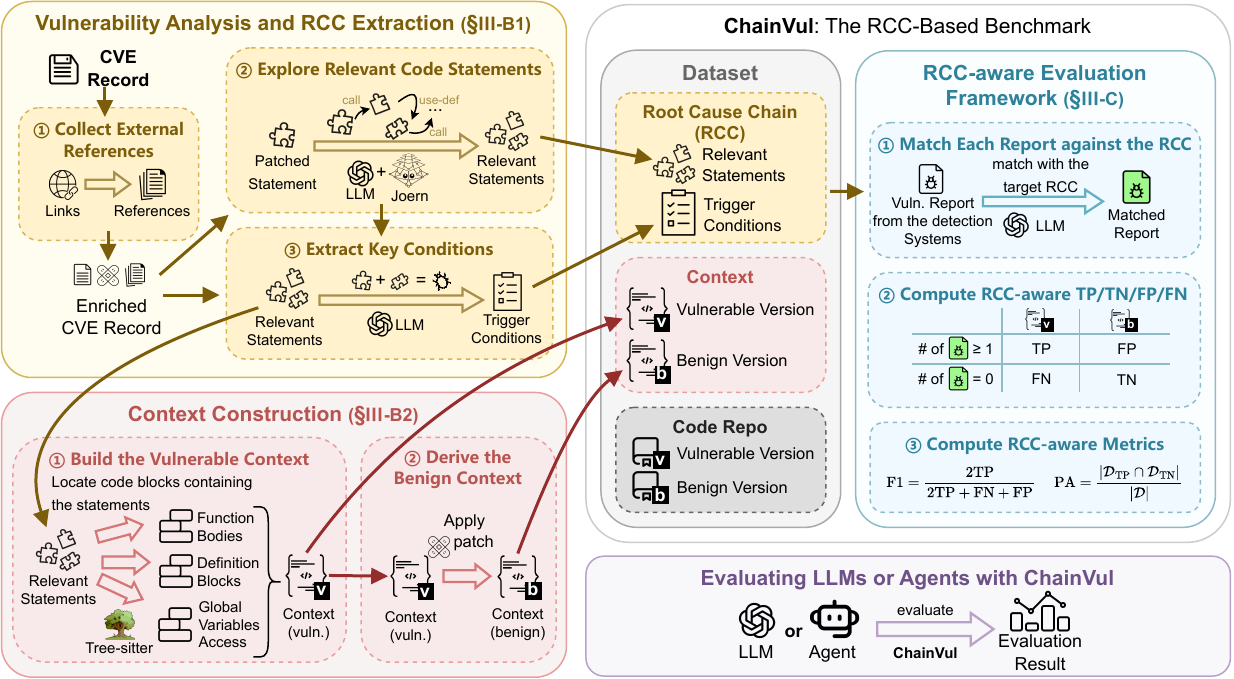}
    \caption{Overview of the RCC-based Benchmark: ChainVul}
    \label{fig:overview}
\end{figure*}

\section{RCC-based Benchmark Construction Approach}
As illustrated in the previous section, coarse-grained code contexts and evaluation metrics in existing benchmarks can lead to imprecise assessment results. To bridge the gap, we propose a fine-grained approach for constructing an LLM-oriented vulnerability benchmark.
The core of it is the \textbf{Root Cause Chain (RCC)}, which serves as an abstract representation of the execution paths that trigger a vulnerability. 

Fig.~\ref{fig:overview} illustrates the overall workflow of our RCC-based benchmark, including both its construction and its application to evaluating LLMs and agents.
The benchmark comprises an \textit{RCC-based dataset} and an \textit{RCC-aware evaluation framework}, providing precise and complete code contexts for vulnerability localization and path-aware report assessment respectively.
For dataset construction, we first analyze each vulnerability and extract its corresponding RCC. The extracted RCC is then used to identify and construct vulnerability-related code contexts, which serve as detector inputs.
For evaluation, the generated reports are assessed by our RCC-aware evaluation framework, which computes RCC-aware metrics to measure the vulnerability detection performance.

In the remainder of this section, we first present formal definitions of the RCC-related concepts, and then elaborate on the detailed benchmark construction methodology and its implementation, called  \textit{{ChainVul}}.

\subsection{Definitions of RCC-related Concepts}

For a specific vulnerability in a real-world repository, the concrete execution paths that can trigger it may be numerous or even infinite, so that using a concrete path to represent the root cause of a vulnerability is improper.
To address this, we employ the \textbf{Root Cause Chain (RCC)} as an abstract representation that characterizes the complete set of vulnerability-triggering paths without explicit enumeration.
\begin{definition}[Root Cause Chain (RCC)]\label{def:rcc}
~\\
A Root Cause Chain of a vulnerability is formally defined as a pair $\rcc = (\Lambda, \Phi)$, where
\begin{itemize}
    \item {$\Lambda$ is a set of code statements consisting of two disjoint parts:
    \begin{itemize}
        \item \emph{key statement set} $\Lambda_K$: statements that are indispensable for triggering the vulnerability;
        \item \emph{supporting statement set} $\Lambda_S$: statements that are not indispensable for triggering the vulnerability, but merely help justify the reachability or data flow between key statements;
    \end{itemize}
    }
    \item $\Phi$ is the set of vulnerability-triggering key conditions over $\Lambda$ that characterize the \textit{source}, \textit{propagation}, and \textit{sink} of the vulnerability-triggering execution paths.
\end{itemize}
\end{definition}

\par\noindent In the motivating example (Fig.~\ref{fig:motivating_example}), the RCC is
\[
\rcc = (\{\lan_5, \lan_8, \lan_9, \lan_{12}, \lam_6, \lam_7\}, \{\phi_1, \phi_2, \phi_3\}),
\]
where $\lan_i$ and $\lam_i$ denote the statements at line $i$ in the files \texttt{nntp.c} and \texttt{memory.c}, respectively. Here $\Lambda_K = \{\lan_9, \lan_{12}, \lam_7\}$ are the key statements that directly perform a key step of the vulnerability, i.e., produce or dereference the null pointer, while $\Lambda_S = \{\lan_5, \lan_8, \lam_6\}$ are supporting statements that only establish the data-flow and caller-callee links between key statements. The triggering condition $\phi_2$ is anchored at the supporting statement $\lan_5$, since $\Phi$ ranges over $\Lambda$ rather than over $\Lambda_K$.\par $\phi_1$, $\phi_2$, and $\phi_3$ denote the three triggering conditions of this RCC. These conditions are as follows:

\begin{itemize}
    \item $\phi_1$: At $\lam_7$, a null pointer is returned (source condition).
    \item $\phi_2$: At $\lan_5$, {\fcmsg} is assigned the return value of {\funcm} (propagation condition).
    \item $\phi_3$: At $\lan_9$ or $\lan_{12}$, {\fcmsg} is dereferenced (sink condition).
\end{itemize}

Based on the extraction of such RCCs, we can construct the \textbf{RCC-based Benchmark}, which consists of two core components: an \textit{RCC-based Dataset} that provides complete and fine-grained descriptions of vulnerabilities, and an \textit{RCC-aware Evaluation Framework} that precisely evaluates the detection results for a target vulnerability. Now we formally define \textit{RCC-based Dataset} as follows.

\begin{definition}[RCC-based Dataset]\label{def:dataset}
~\\
    The RCC-based dataset is formally defined as a set of vulnerability samples:
   $\Dd = \{(\rcc, \hat{\Pp}, \hat{\Kk})\}$, where each sample corresponds to a distinct vulnerability. For a given vulnerability $v$, 
   \begin{itemize}
       \item $\rcc$ is the root cause chain of vulnerability $v$.
       \item $\hat{\Pp} = (\Pp, \Pp')$ denotes the pair of complete repository snapshots before and after applying the patch to fix the target vulnerability $v$, respectively.
       \item $\hat{\Kk} = (\Kk, \Kk')$ denotes the vulnerability-related \textit{code contexts} in repositories, 
       where $\Kk \subseteq \Pp$ and $\Kk' \subseteq \Pp'$. 
       \end{itemize}
\end{definition}

\subsection{RCC-based Dataset Construction}

Given a vulnerability case, which is a CVE record in this work, this section describes how we analyze the case and transform it into a sample in our RCC-based dataset.

\subsubsection{\textbf{Vulnerability Analysis and RCC Extraction}}
\label{sec:rcc-construction}

To construct an RCC, we first analyze the vulnerability and identify the evidence required to explain its manifestation. The extraction process consists of the following three steps. 

\textbf{$\bigstar$ Collect external references.} We first collect external information associated with the vulnerability. Specifically, we crawl all reference links recorded in the CVE, including discussions, announcements, blogs, and other related webpages, which often provide valuable insights that cannot be directly inferred from the source code alone. We further use the GitHub API (Application Programming Interface) to recursively retrieve all issues, commits, and pull requests relevant to the vulnerability.
The collected content is then summarized by an LLM, which removes irrelevant information and produces a compact description that is used in subsequent steps.
 
\textbf{$\bigstar$ Explore relevant code statements.}
The second step aims to identify all code statements that provide essential evidence for understanding the triggering process of the vulnerability. 
The extraction process is implemented as an LLM-guided static analysis workflow that integrates the static analysis tool Joern~\cite{joern}. Starting from the initial patched statements, the LLM iteratively analyzes the vulnerability context and generates Joern queries to guide further exploration. Joern then executes these queries to retrieve additional relevant statements and their dependencies. This iterative process continues until sufficient evidence has been collected to characterize the vulnerability.
To ensure both completeness and reliability of the extracted results, we adopt two exploration strategies: \emph{name-level semantic completeness}, which ensures sufficient semantic evidence for understanding the vulnerability, and \emph{code-content consistency}, which ensures that all extracted statements faithfully correspond to the original source code. The exploration terminates when both conditions are satisfied.

\textit{Name-Level Semantic Completeness Strategy.}
A key challenge in statement exploration is determining whether the collected statements provide sufficient semantic evidence for understanding the vulnerability. For project-specific APIs, stopping at a call site may leave behavior unclear, whereas recursively expanding all callees can introduce substantial noise. Thus, a key API is considered semantically complete only if its name clearly indicates the relevant behavior or its implementation is included in another relevant statement.
For the example in Fig.~\ref{fig:motivating_example}, the vulnerability-related function {\funcn} invokes {\funcm}, which further calls the standard library API \texttt{calloc()}. While the semantics of \texttt{calloc()} are generally self-evident, understanding the precise behavior of {\funcm} may require inspecting its implementation. This strategy ensures that RCCs contain sufficient semantic information while avoiding unnecessary exploration noise.

\textit{Code-Content Consistency Strategy.}
Another challenge arises from LLM hallucination, where the exploration process may generate statements that do not exist in the source code. To improve the reliability of RCC extraction, each extracted statement is required to include both its line number and code content. We then verify consistency by checking whether the extracted content exactly matches the original source code after whitespace normalization. Statements failing this verification are discarded, ensuring that all remaining statements faithfully correspond to real source-code artifacts.

\textbf{$\bigstar$ Extract key conditions.}
Based on the vulnerability-relevant code statements extraction, we further identify the \textit{key conditions} that characterize the target vulnerability. 
A vulnerability may be triggered through multiple execution paths, where each path contains a sequence of \textit{conditions} describing step-wise execution constraints. However, concrete paths often include numerous \textit{irrelevant conditions} that are path-specific and do not determine whether the path corresponds to the vulnerability. Therefore, our goal is to extract all \emph{key conditions} that define the vulnerability while filtering out path-specific details that are not essential for vulnerability characterization.
For example, consider the motivating example in Fig.~\ref{fig:motivating_example}, where three key conditions $\phi_1$, $\phi_2$, and $\phi_3$ have been discussed before. In addition, there exists a condition $\phi'$ (i.e., at $\lan_8$, \texttt{rc} equals zero), which only captures a path-specific constraint along execution path $p_1$. However, another execution path $p_2$ does not satisfy $\phi'$, yet still leads to the dereference of {\fcmsg} at $\lan_{12}$. Thus, $\phi'$ is an \textit{irrelevant condition}.
This example shows that key conditions should characterize the vulnerability at the semantic level rather than overfitting to specific execution paths.

To extract key conditions, we first filter out vulnerability-irrelevant conditions. The remaining relevant conditions are then organized using a \emph{source-propagation-sink} structure, which captures how a vulnerability is triggered: where the focal variable originates, how it propagates through the program, and where it eventually triggers the vulnerability. In the motivating example, $\phi_1$, $\phi_2$, and $\phi_3$ correspond to the source, propagation, and sink conditions, respectively.
Notably, the source and sink are not restricted to single statements; instead, each may correspond to a set of candidate statements. In our implementation, we leverage an LLM to extract such structured key conditions, and carefully design prompts to avoid overly coarse or overly fine-grained abstractions.

\subsubsection{\textbf{Context Construction}}
\label{sec:context-extraction}

After extracting the RCC, we obtain a set of relevant statements and triggering conditions. We construct context solely based on these relevant statements. In particular, we first establish the context for the vulnerable version and subsequently derive the benign context.

\textbf{$\bigstar$ Build the vulnerable context.}
For the vulnerable version, we locate all code blocks that contain the relevant statements using tree-sitter\cite{tree-sitter}. These blocks include function bodies, definition blocks of structs, types, and macros, as well as code blocks that access or manipulate global variables. The union of these code blocks constitutes the context of the vulnerable version.

\textbf{$\bigstar$ Derive the benign context.}
Based on the extracted vulnerable context, we further derive the benign context. Since the benign version differs from the vulnerable version only by the corresponding fixing patch, we obtain the benign context by applying the patch to the extracted vulnerable context.

\subsection{RCC-aware Evaluation Framework}
\label{sec:metric}
\label{sec:vuln-match}

With the RCC-based dataset as input to the LLM-based vulnerability detection systems, we further introduce an RCC-aware evaluation framework for fine-grained performance assessment. 
Notably, the RCC-based dataset (cf. Definition~\ref{def:dataset}) is inherently a paired dataset, where each sample consists of a vulnerable repository snapshot $\Pp$ and its corresponding fixed version $\Pp'$ after patch application. Given this paired nature, the evaluation framework requires the detection system to generate reports for both $\Pp$ and $\Pp'$ simultaneously. The framework then takes these two sets of reports as input and computes evaluation metrics guided by the RCC of the target vulnerability, following three steps.

\textbf{$\bigstar$ Match Each Report against the RCC.}
We match each report against the RCC of the target vulnerability. Recall that an RCC consists of code statements $\Lambda$ and triggering conditions $\Phi$. A report typically describes a vulnerability through multiple behavior descriptions, each of which may cite specific code fragments as supporting evidence. To determine whether a report matches the target RCC, we perform the following two checks.

{First, we \emph{check the key statements cited by each behavior} against $\Lambda_K$. Each behavior in the report typically refers to one or more specific key statements in the code. We perform an inclusion check to determine whether each of the key statements has been cited at least once by the behaviors. Because key statements represent the necessary operations for triggering the vulnerability, the report is considered not matched to the target vulnerability if a key statement is not cited anywhere in the report.}

Second, we \emph{check whether each behavior satisfies the key conditions} in $\Phi$. Each condition in $\Phi$ represents a key logical constraint that must hold along the vulnerability-triggering paths. We verify that the behaviors described in the report collectively conform to these constraints. If any condition in $\Phi$ is not satisfied by the corresponding behavior(s), the report is considered not matched to the target vulnerability.

A report is matched to the target vulnerability only if it passes both the {key-statement inclusion} check and the condition-satisfaction check.
In our implementation, we use an LLM-as-judge to perform the two checks above.

\textbf{$\bigstar$ Compute RCC-aware TP/TN/FP/FN.}
After RCC matching, we obtain all matched reports. In the subsequent step, we focus only on these matched reports. 
For a given sample $d = (\rcc_d, \hat{\Pp_d} = (\Pp_d, \Pp_d'), \hat{\Kk_d}) \in \Dd$ from the RCC-based dataset,
we denote $\Rr_d$ and $\Rr_d'$ as the $\rcc_d$-matched reports set generated from the vulnerable program $\Pp_d$ and the benign program $\Pp_d'$, respectively. Note that, even for the fixed (benign) version, the system may still incorrectly report the presence of the target vulnerability; such reports are also included in the matched set.

Based on the paired reports $\Rr_d$ and $\Rr_d'$ between the vulnerable and benign versions, we classify each sample $d$ into four subsets ($\DTP$, $\DFP$, $\DFN$, $\DTN$) of $\Dd$, which serve as the foundation for computing evaluation metrics. {Here, we count \textit{true positive} (TP) samples, which have at least one matched report in the vulnerable version; \textit{false negative} (FN) samples, which have no matched report in the vulnerable version; \textit{true negative} (TN) samples, which have no matched report in the benign version; and \textit{false positive} (FP) samples, which have at least one matched report in the benign version. Accordingly, $\DTP$, $\DFN$, $\DTN$, and $\DFP$ denote the sets of TP, FN, TN, and FP samples, respectively.} Formally, $\DTP = \{d \in \Dd \mid |\Rr_d| \geq 1\}$,
$\DTN = \{d \in \Dd \mid |\Rr_d'| = 0\}$; $\DFP = \{d \in \Dd \mid |\Rr_d'| \geq 1\}$; $\DFN = \{d \in \Dd \mid |\Rr_d| = 0\}$.
Finally, the four evaluation counts $\TP$, $\FP$, $\FN$, and $\TN$ are obtained as the sizes of $\DTP$, $\DFP$, $\DFN$, and $\DTN$, respectively, i.e., $\TP = |\DTP|$, $\FP = |\DFP|$, $\FN = |\DFN|$, and $\TN = |\DTN|$.

\textbf{$\bigstar$ Compute RCC-aware Metrics.} With the computed counts of TP, FP, TN, and FN, we use two RCC-aware metrics to evaluate detection performance.

\begin{itemize}
    \item \textbf{F1-Score (F1):} F1 summarizes overall detection quality by balancing precision and recall:
    \[F1 = \frac{2P \cdot R}{P + R}\]
    where \(P = TP / (TP + FP)\) and \(R = TP / (TP + FN)\) denote precision and recall, respectively.
    \item \textbf{Paired Accuracy (PA):} PA measures whether a detection system correctly distinguishes a vulnerable sample from its paired fixed counterpart. 
    For a given detection system,  a sample $d$ is considered \textit{paired-correct} if the system reports the vulnerability in the vulnerable program $\Pp_d$ (i.e., $|\Rr_d| \ge 1$) while producing no report for the fixed version $\Pp_d'$ (i.e., $|\Rr_d'| = 0$).
    In other words, a sample $d$ is \textit{paired-correct} iff $d \in \DTP \cap \DTN$. Note that since $\DTP$ and $\DTN$ are not mutually exclusive, their intersection exactly captures the set of samples where both conditions hold. Thus, PA is computed as:
    $$PA = \frac{|\Dd_\TP \cap \Dd_\TN|}{|\Dd|}$$ 
\end{itemize}

\subsection{Benchmark Implementation}
\label{sec:implementation}

Following the construction procedure described above, we develop a real-world vulnerability benchmark, \textbf{ChainVul}. 
To ensure consistent data sources and standardized repository formats, while maintaining the scalability of Joern-based static analysis, we select CVE samples from MegaVul~\cite{MegaVul} whose corresponding repositories are hosted on GitHub and whose repository sizes are smaller than 200~Megabytes (MB). This filtering process yields 3,299 samples. 
We then crawl the reference links associated with these samples and retain only those for which all referenced pages remain accessible, resulting in 2,129 samples.
These 2,129 samples serve as the candidate pool for constructing the final RCC dataset.

{RCC extraction requires 3.6 exploration rounds on average, and 83.1\% of samples converge within 6 rounds. Exploration terminates when every extracted statement satisfies both the name-level semantic completeness and code-content consistency criteria; samples failing due to timeouts or incomplete responses are retried up to three times before being discarded.}

Then, we construct a real-world dataset containing 1,000 RCC-annotated samples from this candidate pool via a random sampling strategy. To preserve the distribution of vulnerability in MegaVul, we iteratively pick samples, apply the RCC extraction pipeline, and validate the results. Invalid cases are discarded, such as LLM timeouts or responses with missing or incomplete information, resulting in a final benchmark ChainVul of 1,000 samples. {The filtering criteria are category-agnostic. The CWE distribution before and after filtering remains consistent. Among the ten Layer-1 CWE categories, six deviate by less than $1\sigma$, where $\sigma$ denotes the standard deviation of a CWE category's share under random sampling.}

\section{Evaluation of ChainVul}
\label{sec:eval-bench}

To evaluate the effectiveness of ChainVul, we investigate the following three research questions.

\begin{itemize}
    \RQGroup{1}
    \item \RQ{bench-rcc} {Accuracy of RCC} Are the extracted RCCs complete and accurate?
    \item \RQ{bench-context} {Effectiveness of RCC-based Context} Does the extracted context effectively expand the original patch?
    \item \RQ{bench-eval} {Reliability of RCC-aware Framework} Does our evaluation framework achieve higher assessment reliability?
\end{itemize}

\subsection{Setup}
\noindent\textbf{Models.}
\label{sec:models}
We use four LLMs: GPT-5.5~\cite{openai}, DeepSeek-V4-Flash (non-thinking), DeepSeek-V4-Flash (thinking), and Qwen3.5-27B. This set covers a strong closed-source baseline, two modes of the same model family to isolate the effect of explicit reasoning, and a smaller open-source baseline. We set the temperature to 0 for all runs to eliminate sampling variance.
For \rqref{bench-eval}, reports from all four models are included so that the matching metric is tested against outputs with heterogeneous quality and reasoning styles. For RCC-based matching, we use GPT-5.5 as the judge model throughout to keep the judging protocol fixed across experiments.

\noindent\textbf{Evaluation subsets and annotation protocol.}
\rqref{bench-rcc} and \rqref{bench-eval} require manual verification. For \rqref{bench-rcc}, we inspect whether each RCC contains the {complete key-statement set} and the correct triggering conditions. For \rqref{bench-eval}, we manually determine whether each generated report matches the target vulnerability described by the RCC, which serves as the ground truth for measuring matching accuracy. Because both tasks require expert inspection of vulnerability semantics and cross-checking model outputs, we conduct these analyses on \bench{100}, a 100-sample subset drawn uniformly at random from ChainVul. {Each annotator had access to the CVE description, the fixing patch, and the pre-patch repository, but not to the automatically extracted RCC. They first marked the key statements required to trigger the vulnerability by analyzing value propagation and simulating execution, then cross-checked the annotations: every patched statement had to be covered or explained as irrelevant, and the three independent annotations had to agree. They were then shown the automatically extracted RCC and judged whether its triggering conditions were correct. Disagreements were resolved by discussion, and remaining conflicts by majority voting.}

\subsection{\rqfullref{bench-rcc}}
We first validate the accuracy of the RCC used in our main experiments (extracted with GPT-5.3-codex), then assess the sensitivity to extraction model. To accurately capture the root cause of a vulnerability, an RCC should cover the key statements and explain their triggering conditions correctly.

\par\noindent For each RCC in \bench{100}, we manually identify all key statements as the ground truth and compare them with the automatically extracted key-statement set. The key-statement coverage of an RCC is defined as
\[
\eta=\frac{|\Lambda_K \cap \Lambda_K^\text{GT}|}{|\Lambda_K^\text{GT}|}
\]
where $\Lambda_K$ denotes the key-statement set in the RCC, and $\Lambda_K^\text{GT}$ denotes the manually identified key-statement set, which serves as the ground truth.
Across the 100 inspected samples, the average key-statement coverage reaches 87.8\%.\par

We further manually inspect the triggering conditions of each RCC to verify whether they correctly characterize how the target vulnerability is triggered. Among the 100 inspected samples, 98 RCCs are correct, yielding an accuracy of 98.0\%.

{After validating the accuracy of RCCs extracted with GPT-5.3-codex, we assess RCC extraction's sensitivity to the extraction model. We also extract RCCs for the same 100 samples with two popular models, DeepSeek-V4.1-Flash and Claude-Sonnet-4.6. Sonnet refuses to produce RCCs for 7 CVEs it deems incomplete, whereas DeepSeek completes all 100. We therefore compare the statement sets extracted by the three models on the 93 commonly completed samples. We analyze the consistency of the RCCs extracted by the three models from both statement-level and semantic-level. From the statement-level, we measure the agreement of the key statements extracted by the three models, and the pairwise F1 scores are 70.6\%, 74.4\%, and 68.4\%. From the semantic-level, we compare semantic natural-language descriptions abstracted from concrete RCCs. Each RCC pair is independently annotated by two authors, who then jointly discuss their assessments and rigorously evaluate the consistency. The proportions of consistent samples are 93.6\%, 94.6\%, and 95.7\%. (The values above are reported in the order of GPT-DeepSeek, DeepSeek-Claude, and Claude-GPT.) This suggests that RCC extraction is consistent and high-quality rather than an artifact of a particular model.}

\begin{result}
    {The RCCs in ChainVul faithfully capture vulnerability root causes, with 87.8\% key-statement coverage and 98.0\% triggering-condition accuracy. The RCC extraction method is reusable across various modern models.}
\end{result}

\subsection{\rqfullref{bench-context}}

RCC-based context extraction is designed to construct a context containing all the key statements relevant to the vulnerability. However, do these key statements commonly appear away from the patched function? In other words, is RCC-based context extraction more effective than the patch-based extraction? Here, we compare patch-based and RCC-based contexts on ChainVul from the perspectives of intra- and inter-procedural distributions.

\begin{figure}[ht]
\centering
\includegraphics[width=\linewidth]{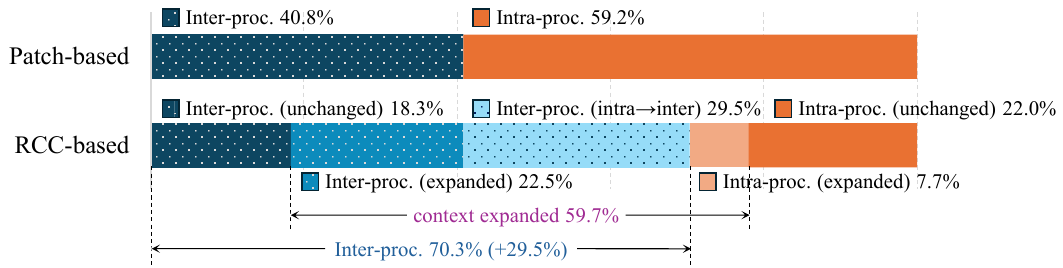}
\caption{Comparison of patch-based and RCC-based contexts}
\label{fig:context-comparison}
\end{figure}

Fig.~\ref{fig:context-comparison} presents the detailed comparison. In the patch-based setting, 40.8\% of the samples are inter-procedural, whereas this proportion increases to 70.3\% under RCC-based context extraction. Notably, 29.5\% of the samples shift from intra-procedural to inter-procedural, indicating that patch-based contexts fail to capture inter-procedural evidence in these cases. In addition, some samples exhibit context expansion without such a shift (22.5\% inter-procedural samples and 7.7\% intra-procedural samples expanded). Overall, the context of 59.7\% of the samples is expanded by RCC-based extraction. These results suggest that traditional patch-based context omits a substantial portion of critical vulnerability evidence, which can be effectively recovered by our RCC-based approach.

\begin{figure}[ht]
\centering
\includegraphics[width=\linewidth]{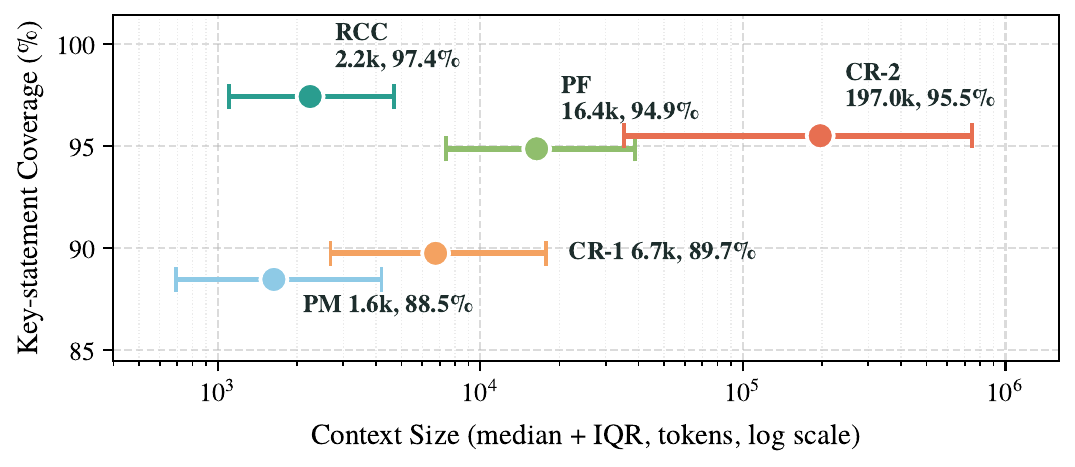}
\caption{{Trade-off between context size and key-statement coverage for different context construction methods.}}
\label{fig:context-static}
\end{figure}

{We further compare RCC-based context with three static-analysis-based constructions: patched methods (PM), patched files (PF), and call-relation expansion (CR). For CR-$k$, the context consists of the patched functions together with all their $k$-hop callers and callees; we use both CR-1 and CR-2 as baselines. Fig.~\ref{fig:context-static} shows their trade-off between key-statement coverage and context size (upper-left is better): RCC achieves the highest coverage with a relatively small context, PM uses a slightly smaller context but covers far fewer key statements, and PF and CR improve over PM at the cost of a 3--100$\times$ larger contexts.}

\begin{result}
    RCC-based context extraction substantially expands the available evidence. {Compared with static-analysis-based constructions, RCC-based context achieves the highest 97.4\% key-statement coverage with a relatively short context of 2.2k tokens.}
\end{result}

\subsection{\rqfullref{bench-eval}}
\label{subsec:eval_fidelity}

To evaluate the assessment fidelity of ChainVul, we analyze the correctness of evaluation outcomes against a human-annotated ground truth, where human experts manually verify whether a report correctly aligns with the RCC. We compare two evaluation strategies: \textbf{Binary (Naive) Matching}, which mimics traditional evaluation by blindly accepting any generated report as positive regardless of its underlying rationale, and \textbf{RCC-based Matching}, which uses an LLM to classify a report as positive only if it aligns with the predefined RCC.
We quantify fidelity using \textit{Matching Accuracy}, defined as the proportion of evaluation outcomes that agree with the human ground truth across 400 reports generated by four LLMs. {Concretely, the ground truth is the human-adjudicated label for each report. Treating a matched report as unmatched, or an unmatched report as matched, is counted as one misjudgment; Matching Accuracy is one minus the misjudgment rate.}

\begin{table}[ht]
\centering
\renewcommand{\arraystretch}{1.2}
\caption{Matching Accuracy Comparison}
\label{tab:fidelity_results}
\begin{tabular}{lcc}
\hline
\textbf{Report Generating Model} & \textbf{\begin{tabular}[c]{@{}c@{}}Binary\\ Matching Acc.\end{tabular}} & \textbf{\begin{tabular}[c]{@{}c@{}}RCC-based\\ Matching Acc.\end{tabular}} \\ \hline
GPT-5.5 & 58\% & 78\% (+20\%) \\
DeepSeek-V4-Flash (non-thinking) & 56\% & 66\% (+10\%) \\
DeepSeek-V4-Flash (thinking) & 40\% & 73\% (+33\%) \\
Qwen3.5-27B & 35\% & 73\% (+38\%) \\ \hline
\end{tabular}
\end{table}

As shown in Table \ref{tab:fidelity_results}, ChainVul consistently improves evaluation fidelity across reports generated by any model, raising the accuracy by about 30\%. 
Crucially, traditional binary evaluation severely misleads the assessment of reasoning-focused models (e.g., DeepSeek with thinking and Qwen3.5), yielding low binary accuracy (40\% and 35\%) due to their verbose, noise-heavy outputs. 
While traditional metrics misinterpret this noise as faulty detection, the RCC-based framework successfully captures valid root-cause alignments, leading to substantial fidelity gains (+33\% and +38\%).

{We further quantify how residual matching errors affect downstream measurements. On the \bench{100}, replacing automatic matching with human-annotated matching leaves the ranking of the four models unchanged. This indicates that although the automatic RCC matching strategy introduces small errors in the F1 scores, it has little effect on the ranking, and can therefore be used to quickly assess and rank the vulnerability detection capabilities of models.}

\begin{result}
    Traditional binary evaluation provides low fidelity (35\%-58\%) by ignoring reasoning paths. ChainVul successfully enforces Root Cause Chain alignment, eliminating evaluation overestimation and achieving fidelity improvement up to 38\%.
\end{result}

\section{Benchmarking LLMs and Agents with ChainVul}
\label{sec:eval-llm}

This section focuses on the new insights provided by evaluating LLMs and agents on ChainVul. Since ChainVul provides more accurate context and a more precise evaluation methodology, it enables a more reliable assessment of LLMs and agent systems. We therefore leverage it to explore the following two aspects.
\begin{itemize}
    \RQGroup{2}
    \item \RQ{eval-llm}{Evaluating LLMs on ChainVul} How do LLMs perform in vulnerability detection with a more complete context? 
    \item \RQ{eval-agent}{Evaluating Agents on ChainVul} How do agent systems perform compared to plain LLMs in exploring and utilizing vulnerability-relevant context?
\end{itemize}

\subsection{Setup}
\noindent\textbf{Evaluated systems.}
For \rqref{eval-llm}, we evaluate the same four LLMs introduced in Section~\ref{sec:models}. We use all four models when comparing RCC-based and patch-based contexts, because that comparison is the main cross-model result of this section. For the CWE-specific analysis, we use GPT-5.5 only. This choice keeps the per-CWE measurements stable while avoiding a fourfold increase in evaluation cost over many category-specific subsets; GPT-5.5 is also the strongest model in Fig.~\ref{fig:context-detect}, making it the most suitable probe for measuring how much additional context can be exploited.

 For \rqref{eval-agent}, we evaluate two agent frameworks, mini-SWE-agent~\cite{swe-agent} and OpenCode~\cite{opencode}, each paired with two backbone models, GPT-5.5 and DeepSeek-V4-Flash (non-thinking). These two backbones represent the strongest and weakest plain-LLM performers in our evaluation, allowing us to test how agentic support benefits models. Both agent frameworks are run under their default configurations so that the comparison reflects their out-of-the-box behavior rather than task-specific tuning.

\noindent\textbf{Evaluation dataset.}
Each ChainVul sample contains both a vulnerable version and a patched version. We evaluate both versions, under both patch-based and RCC-based context settings, for four standalone LLMs and four agent configurations. This results in $2\times 2\times (4+4)=32$ evaluations per sample. To keep this design tractable while preserving diversity, we construct \bench{300}, a uniformly random 300-sample subset of ChainVul, and use it for most experiments in this section. The only exception is the CWE-specific analysis, which uses the full benchmark so that each category-specific subset remains sufficiently large for comparison. {Both \bench{100} and \bench{300} are representative of the full benchmark: every top-15 CWE share lies within the $\pm2\sigma$ sampling interval, where $\sigma$ is the standard deviation of a CWE share under random sampling, and the Pearson correlation of CWE-share vectors is 0.986 for \bench{300} and 0.909 for \bench{100}.}

\subsection{\rqfullref{eval-llm}}
This subsection compares LLM performance under RCC-based and baseline patch-based contexts to measure the effect of context completion.

The context in ChainVul is constructed with RCCs (Section~\ref{sec:context-extraction}), which we denote as \textbf{RCC-based context}. As a baseline, we use all changed code blocks (i.e., function bodies, definition blocks, and global variable accesses), denoted as \textbf{patch-based context}. Based on the relationship between these two contexts, we partition samples in \bench{300} into two groups: \textbf{context-expanded samples (CES)}, where RCC-based context is larger than patch-based context, and \textbf{context-unchanged samples (CUS)}, where both contexts are identical.

\begin{figure}[htbp]
    \centering
    \raggedright\footnotesize\textit{Notes: M\#G = GPT-5.5, M\#DT = DeepSeek-V4-Flash (thinking), \\\hphantom{Notes: }M\#D = DeepSeek-V4-Flash (non-thinking), M\#Q = Qwen3.5-27B.}
    \begin{subfigure}[b]{0.49\linewidth}
        \centering
        \includegraphics[width=\linewidth]{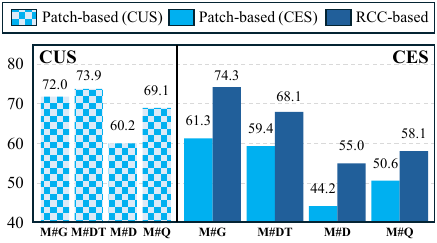}
        \caption{F1-Score (\%) by Model}
        \label{fig:context-detect-a}
    \end{subfigure}
    \hfill  
    \begin{subfigure}[b]{0.49\linewidth}
        \centering
        \includegraphics[width=\linewidth]{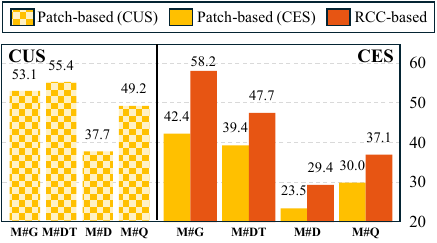}
        \caption{Paired Accuracy (\%) by Model}
        \label{fig:context-detect-b}
    \end{subfigure}
    \caption{Detection Performance under CUS and CES}
    \label{fig:context-detect}
\end{figure}

Fig.~\ref{fig:context-detect} reports results on CES and CUS separately. We draw two main observations.
\textbf{(1) CUS is consistently easier than CES.} For nearly all models, vulnerability detection performance on CUS is higher than on CES (with RCC-based context) by 5.2--11.0\% in F1-score and 7.7--12.1\% in paired accuracy. CUS keeps the evidence largely local, whereas CES often requires reasoning over non-local evidence along longer execution paths. GPT-5.5 is the only exception, showing similar performance on CES and CUS.
\textbf{(2) RCC-revealed evidence improves detection.} On CES, RCC-based context consistently outperforms patch-based context by 7.5--13.0\% in F1 and 5.9--15.8\% in paired accuracy. These gains indicate  RCC-expanded code contains actionable vulnerability evidence missing from the patch-only context.

\begin{finding}
While non-expanded vulnerability samples are inherently more accessible for LLM detection (improving F1-score by 5.2--11.0\%), RCC-based context provides superior precision for context-expanded samples (CES), delivering a substantial 7.5--13.0\% F1-score advancement.
\end{finding}

As context expansion works well to increase the detection capability of LLM-based systems, we next analyze which vulnerability categories benefit most from richer context.
To answer this question, we partition all samples in ChainVul into subsets according to their CWE IDs, and evaluate the context dependence of each subset using three indicators: CER, $\Delta \text{F1}$, and $\Delta \text{PA}$. CER (Context-Expansion Rate) denotes the proportion of samples in a subset whose context is expanded by RCC. $\Delta \text{F1}$ and $\Delta \text{PA}$ denote the gains in F1 score and paired accuracy, respectively, achieved by RCC-based context over patch-based context. Both $\Delta \text{F1}$ and $\Delta \text{PA}$ are computed based on the detection results of the best-performing model (GPT-5.5).

\begin{figure}[h]
    \centering
    \includegraphics[width=\linewidth]{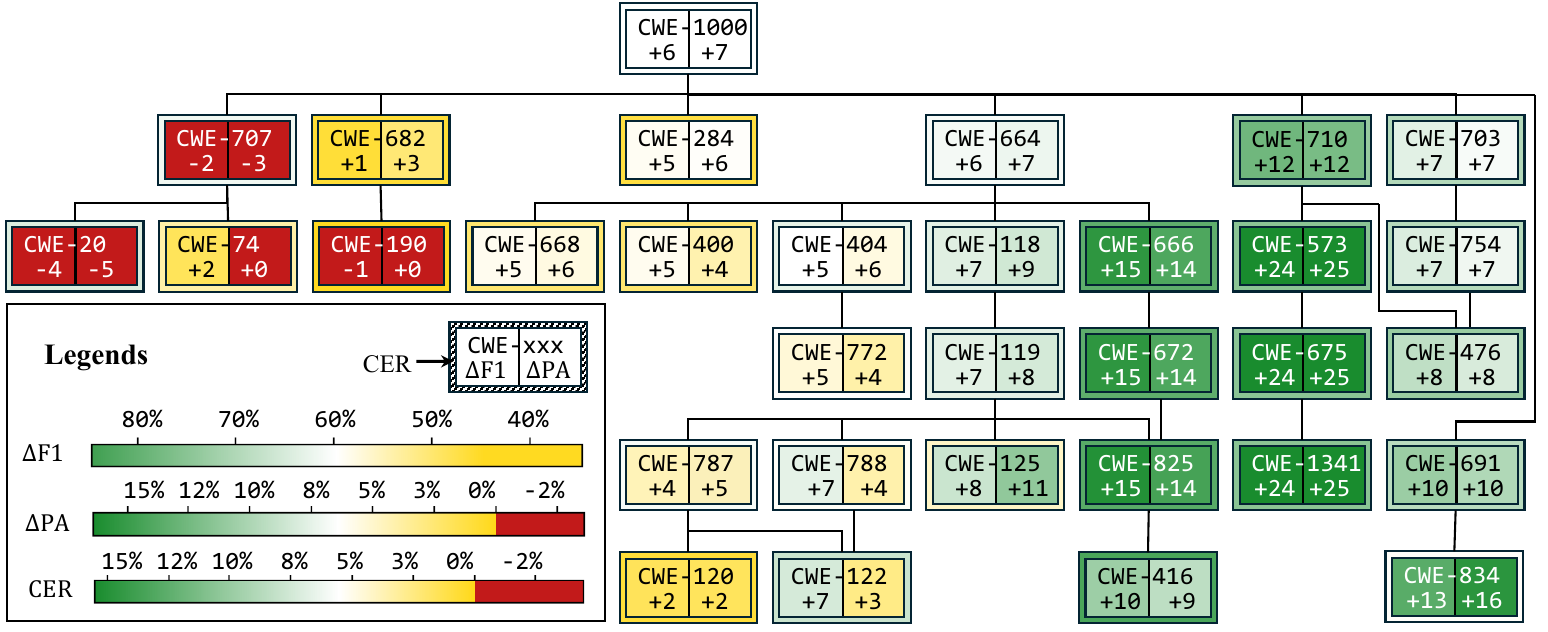}
    \caption{Analysis of Context Dependence across CWEs}
    \label{fig:cwe}
    \vspace{-12pt}
\end{figure}

\begin{figure*}[!t]
    \centering
    \raggedright\footnotesize\textit{Notes: Standard = visible of the entire repository, Context-guided = only visible of the vulnerability-relevant context, MSA = mini-SWE-agent}\\
    \begin{minipage}[t]{0.55\textwidth}
    \centering
    \includegraphics[width=.85\linewidth]{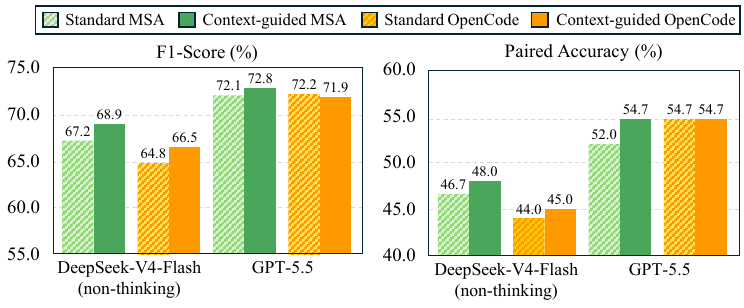}
    \caption{Detection Performance of Standard \& Context-guided Agents}
    \label{fig:eval-agent-explore}
    \end{minipage}\hfill
    \begin{minipage}[t]{0.44\textwidth}
    \centering
    \includegraphics[width=\linewidth]{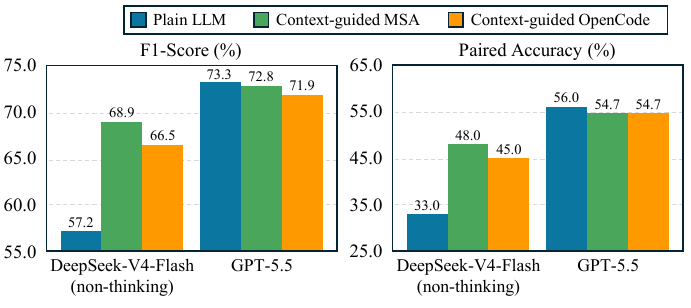}
    \caption{Detection Performance of Plain-LLM and Agents}
    \label{fig:eval-agent-util}
    \end{minipage}
    \vspace{-0.2cm}
\end{figure*}

Fig.~\ref{fig:cwe} visualizes these indicators for each CWE as a double-bordered node: the outer band denotes CER, the left half denotes $\Delta \text{F1}$, and the right half denotes $\Delta \text{PA}$. White marks values near the mean, green above-average context dependence, and yellow below-average dependence; stronger saturation indicates larger deviations. Negative or near-zero gains suggest that RCC expansion adds little useful evidence or introduces a distraction.

Overall, the three indicators tend to vary consistently across CWEs, as reflected by the coherent color patterns within individual nodes. This agreement suggests that different metrics lead to the same overall conclusion, strengthening the credibility of our findings. Moreover, parent and child CWEs often appear in similar color regions, indicating that related vulnerability families tend to exhibit comparable context requirements. Finally, representative case analysis further reveals that context-dependent and context-independent vulnerabilities focus on distinct code-level operations, providing additional qualitative support for the observed patterns.

To better illustrate these distinctions, we further separate CWEs into context-independent and context-dependent categories.
For \textit{context-independent categories}, the vulnerability is typically determined by a single core operation, with limited dependence on surrounding context. For example, CWE-682 (Incorrect Calculation) is primarily characterized by the erroneous computation itself. Similarly, CWE-74 (Injection) is often centered on the vulnerable statement that executes or propagates untrusted input.
By contrast, \textit{context-dependent categories} usually require reasoning over a chain of operations. For example, CWE-1341 and CWE-825, which involve improper resource release timing (e.g., double free and use-after-free), require tracking resource allocation and deallocation across a potentially long execution path. Likewise, CWE-834 (Excessive Iteration), including uncontrolled recursion and infinite loops, depends heavily on reasoning about interactions among multiple program components along the loop structure.

\begin{finding}
Context dependence is consistent across metrics and the CWE hierarchy, and is typically higher for vulnerabilities that involve sequences of interdependent operations rather than isolated ones.
\end{finding}

\subsection{\rqfullref{eval-agent}}
As ChainVul provides a more complete code context for evaluation, we further assess agents from two perspectives: their ability to explore vulnerability-relevant context and their ability to effectively utilize such context.

\textbf{Ability to explore context.} 
A key distinction between agents and standalone LLMs is that agents can interact with the entire repository. This raises a natural question: \emph{Can an agent successfully discover all vulnerability-relevant context through exploration?} In principle, this could be evaluated by tracing all regions accessed by the agent and comparing them with the annotated relevant context in ChainVul. However, in practice this is challenging, as agents may inspect code through diverse and non-deterministic interaction paths.
To enable such an evaluation, we instead introduce a comparison between a standard agent and a \emph{context-guided agent}. The \emph{context-guided agent} operates on a masked version of the repository, where only vulnerability-relevant code is exposed and all other lines are replaced with dummy placeholders (\texttt{<masked>}). In addition, the prompt explicitly provides the locations of all visible code regions, ensuring that the agent’s exploration is constrained to the available context.

Based on this experimental design, Fig.~\ref{fig:eval-agent-explore} compares standard and context-guided agents across two frameworks and two backbone models. Surprisingly, the context-guided variant yields only marginal improvements in both F1 score and paired accuracy, indicating that standard agents are already able to recover most vulnerability-relevant context without explicit guidance.

{We also directly examine whether standard agents explore the location of the root cause. We align the explored regions with the key statements $\Lambda_K$ on the vulnerable (before-fix) version and define key-statement coverage (KSC) as the fraction of key statements that an agent has explored. On the human-annotated \bench{100}, KSC is 95.6\% for TP samples and 87.2\% for FN samples. KSC exceeds 85\% for both TP and FN samples, indicating that agents already have a strong ability to recover vulnerability-relevant context. This is consistent with the marginal gains of context-guided agents above.}

\textbf{Ability to utilize context.}
Beyond identifying vulnerability-relevant context, we further investigate whether agents can effectively leverage such context during detection. Specifically, we ask whether a context-guided agent can outperform a plain LLM when both are provided with the same ground-truth context from ChainVul.
To answer this, we feed the annotated context from ChainVul into a plain LLM and compare its performance with a context-guided agent built on the same backbone model.
Fig.~\ref{fig:eval-agent-util} shows that the agent substantially improves performance over the plain LLM for DeepSeek-V4-Flash in non-thinking mode, increasing F1-score by 11.7\% and paired accuracy by 15.0\%.  By contrast, for GPT-5.5, the agent does not yield further gains; instead, it slightly reduces performance, with both metrics dropping by 0.5--1.4\%.
\begin{finding}
    Agents exhibit a stronger-than-expected capability in exploring and identifying relevant code context within repositories, regardless of the underlying LLM used. In contrast, their ability to utilize the same context depends on the underlying model: the agent framework significantly improves performance for weaker, non-thinking LLMs, but provides only marginal gains for stronger models that already effectively exploit contextual information.
\end{finding}

\section{Threats to Validity}
Our study has several threats to validity.
\textbf{First}, ChainVul is built from MegaVul, which mainly covers C/C++ vulnerabilities. Our filtering criteria---GitHub-hosted repositories, size below 200MB, and accessible external references---improve feasibility and reproducibility, but may introduce selection bias. Although the benchmark preserves MegaVul's overall CWE distribution, it may differ in other aspects.
\textbf{Second}, due to evaluation cost, some experiments use sampled subsets rather than the full benchmark: \bench{100} for manual RCC and matching validation, and \bench{300} for most LLM and agent experiments. Although uniformly sampled, these subsets may still introduce bias. {We verified that the CWE distributions of the subsets are consistent with that of the full benchmark.}
\textbf{Third}, RCC-aware evaluation relies on a fixed judge model, which ensures consistency but is not perfectly reliable. To increase result credibility, we  adopt GPT-5.5, a representative state-of-the-art model, as a more capable judge. {We further analyzed the impact of RCC-based matching errors, and the results show that these errors do not significantly affect the experimental conclusions.}
\textbf{In addition}, manually annotated data may still contain errors despite independent annotation and reconciliation.

\section{Conclusion}
We presented \textit{ChainVul}, a root-cause-chain-based benchmark for precise evaluation of LLMs and agents on real-world vulnerability detection. Its core concept, the \emph{Root Cause Chain} (RCC), captures both vulnerability-relevant statements and triggering conditions, enabling fine-grained context construction and RCC-aware assessment. Results show that RCC annotations are highly accurate, RCC-based contexts recover substantial evidence missed by patch-based annotations, and RCC-aware matching improves evaluation fidelity by about 30\%.
We further show that context completeness and utilization are both critical, and that agentic support mainly benefits weaker backbone models.
We believe \textit{ChainVul} provides a practical foundation for more reliable evaluation of LLM-based vulnerability detection and future research on context-aware, reasoning-centric security analysis.

\section{Data Availability}
Our dataset and code, including all prompts, are publicly available~\cite{ourtool}.

\clearpage

\bibliographystyle{IEEEtran}
\bibliography{ref}

\end{document}